\documentclass[a4paper,11pt]{article}
\pdfoutput=1 

\usepackage{jinstpub} 
\usepackage{epsfig}

\usepackage{lineno}
\usepackage{hyperref}

\newcommand\blfootnote[1]{%
  \begingroup
  \renewcommand\thefootnote{}\footnote{#1}%
  \addtocounter{footnote}{-1}%
  \endgroup
}

\title{\boldmath Upgraded Electronics of the ATLAS Hadronic Tile Calorimeter for the High Luminosity LHC}

\author[a]{J. Abdallah,\note{Corresponding author.}}

\affiliation[a]{The University of Texas at Arlington,502 Yates Street, USA}

\emailAdd{Jalal.Abdallah@cern.ch}


\collaboration[c]{on behalf of ATLAS Tile Calorimeter group}

\proceeding{CHEF2019 - CALORIMETRY FOR THE HIGH ENERGY FRONTIER\\
  25-26 November 2019\\
  FUKUOKA, JAPAN}

\abstract{The ATLAS hadronic Tile Calorimeter will undergo major upgrades to the
on- and off-detector\footnote{Refers to a technical cavern located $\sim$60m away from the detector cavern.}
electronics in preparation for the High Luminosity program of the Large Hadron Collider (HL-LHC) in 2026, so that the
system can cope with the HL-LHC increased radiation levels and out-of-time pileup.
The on-detector electronics of the upgraded system will continuously digitize and 
transmit all photo-multiplier signals to the off-detector systems at a 40 MHz rate. 
The off-detector electronics will store the data in pipeline buffers, produce 
digital hadronic tower sums for the ATLAS Level-0 trigger system, and read
out selected events. The modular on-detector electronics feature radiation-tolerant commercial 
off-the-shelf components and redundant design to minimize single points of failure. 
The timing, control and communication interface with the off-detector electronics
is implemented with modern Field Programmable Gate Arrays and high speed fibre 
optic links running up to 9.6 Gbps.
\blfootnote{~}
\blfootnote{\tiny \textsuperscript{\textcopyright} Copyright 2020 CERN for the benefit of the ATLAS Collaboration. \newline
Reproduction of this article or parts of it is allowed as specified in the CC-BY-4.0 license.}.
}

\begin{document}
\maketitle
\flushbottom

\section{Introduction}

The ATLAS Tile Calorimeter~\cite{tile} (TileCal) is the central hadronic calorimeter of the ATLAS experiment~\cite{atlas} at
the Large Hadron Collider (LHC). It is a sampling calorimeter made of scintillating plastic tiles as the active 
material and steel plates as passive absorbers. The detector is composed of three cylinders, a long barrel and two extended barrel cylinders, 
covering a pseudo-rapidity range of $\eta < 1.7 $. Each cylinder is segmented into 64 modules along the azimuth angle $\phi$.
The long barrel cylinder is divided into two readout sections and named LBA and LBC. The extended barrels located on each side of the long barrel are called 
EBA and EBC. Wavelength shifting fibers (WLS) collect the light, generated by the passage of particles in the tiles, and transmit it to photo-multipliers (PMT).
The tiles are grouped in cells of different dimensions depending on the pseudo-rapidity and depth and the WLS fibers of a given cell are coupled to two PMTs for redundancy. Three radial layers A, BC, D are defined 
inside the modules. The obtained segmentation gives rise to projective towers with granularity of $\Delta \eta \times \Delta \phi = 0.1 \times 0.1$ in the first 
two layers A and BC and $\Delta \eta \times \Delta \phi = 0.2 \times 0.1$ in the outer layer D. Two additional special cells and four scintillator slabs (Intermediate TileCal) instrument the region between LB and EB. They are read out by single PMT each.
The modules in the LB are read out using up to 45 PMTs while 32 PMTs are needed for the EB modules. In total, the TileCal has 5182 cells and 9852 PMTs.

The HL-LHC aims for an instantaneous luminosity of up to $7.5 \times 10^{34}\rm{cm}^{-2}\rm{s}^{-1}$ (about five times more compared the the LHC).
This will result in the production of about 200 proton-proton collisions per bunch crossing and a significant increase of the particle flux. TileCal is estimated to receive 
up to about 80 Gy of Total Ionizing Dose (TID) for 4 ab$^{-1}$ of integrated luminosity. The full readout system has to be replaced in order to handle the increase
in the data rate, to be more radiation tolerant for the on-detector electronics and to be compatible with the fully digital ATLAS Trigger/DAQ system for HL-LHC.

\section{The Tile Calorimeter Upgrade}

\begin{figure}[htbp]
\centering 
\includegraphics[width=.9\textwidth]{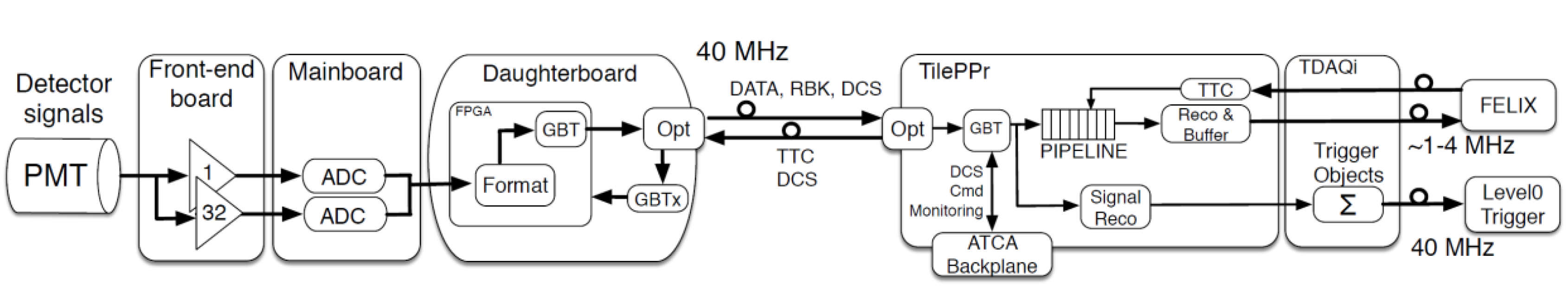}
\caption{\label{fig:readout} Phase-II architecture \cite{tileup}.}
\end{figure}

The Upgrade of the TileCal consists of the full replacement of the on- and off-detector electronics as well as the redesign of the
power distribution systems to implement redundancy and improve reliability.
According to the new requirements for the HL-LHC, electronics parts tolerant to the expected radiation level and capable of sustaining the high
trigger rate up to 1 MHz and a maximum latency of $ 10 \mu s$ were chosen. This required moving the electronics buffers and pipelines to
the off-detector part of the system which means sending the data at the LHC bunch-crossing rate of 40~MHz. 
A schema of the full readout chain for the HL-LHC is shown in Figure~\ref{fig:readout}.

\subsection{Mechanics for the on-board electronics}
\label{sec:mecani}
\begin{figure}[htbp]
\centering 
\begin{tabular}{cc}
\begin{minipage}{0.5\textwidth} \includegraphics[width=.95\textwidth]{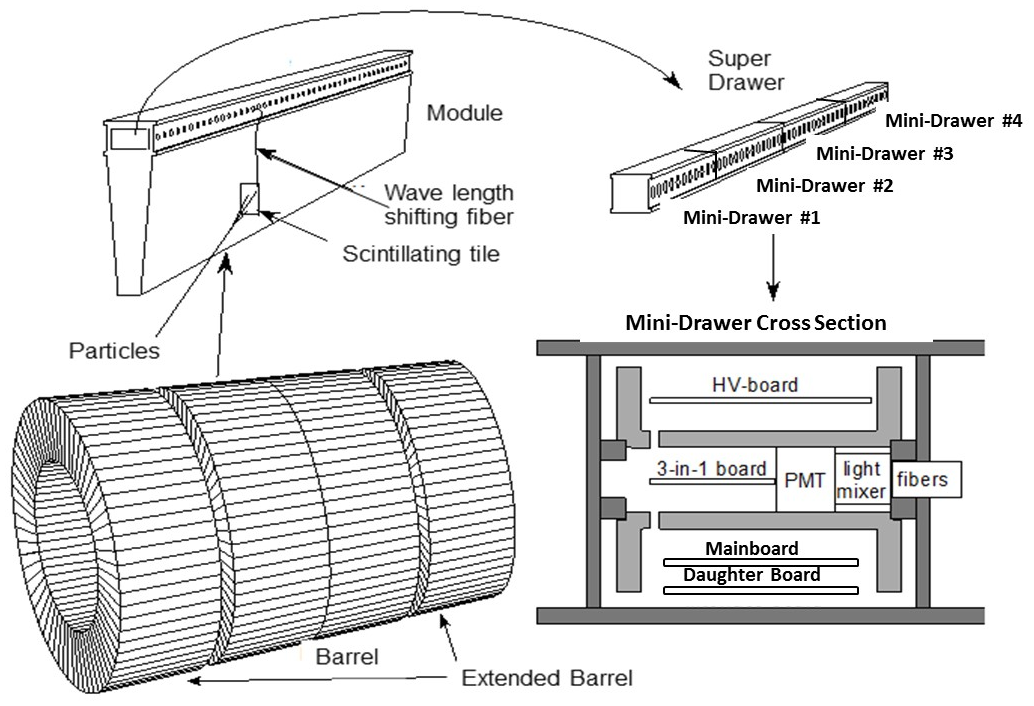} \end{minipage} &  \begin{minipage}{0.5\textwidth} \includegraphics[width=.95\textwidth]{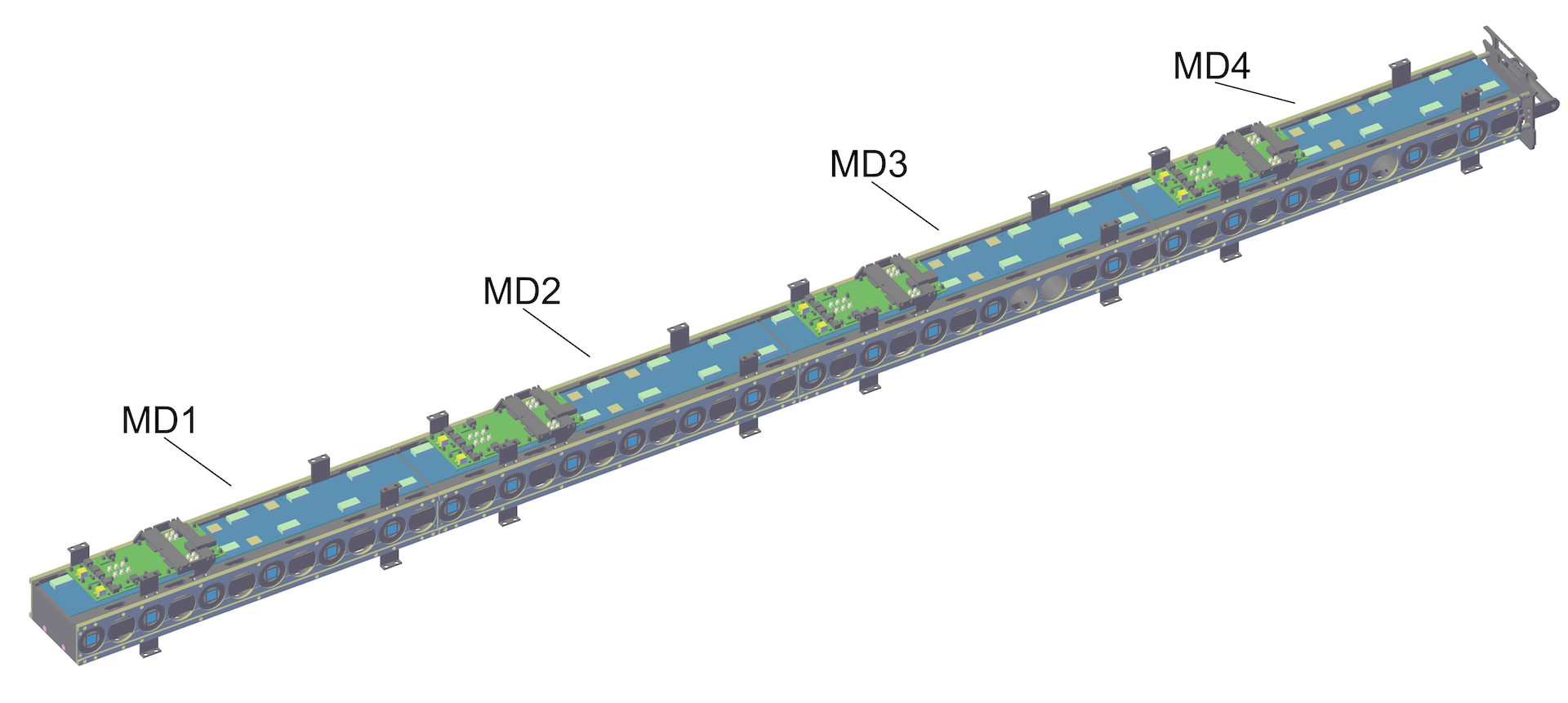} \\ \includegraphics[width=.95\textwidth]{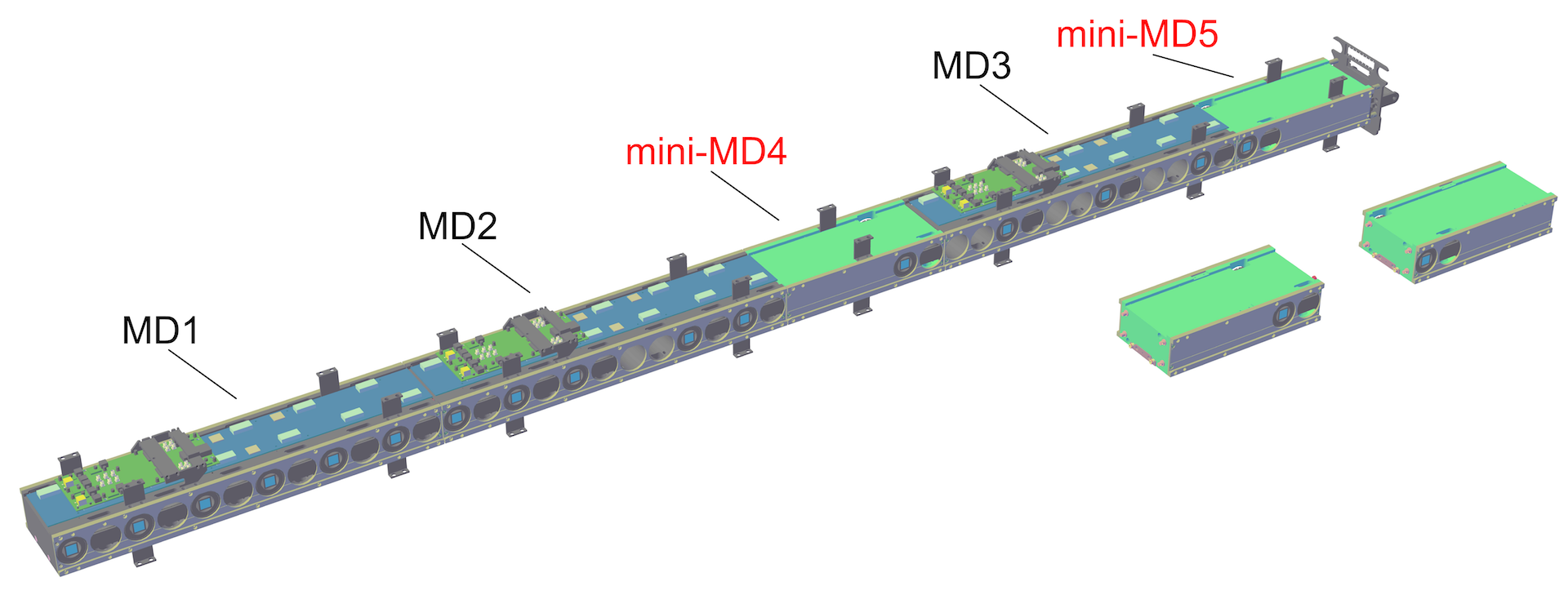} \end{minipage}
\end{tabular}
\caption{\label{fig:meca} Tile calorimeter and on-detector electronics drawers ({\it left}) and the new super-drawer architecture for LB and EB ({\it right})~\cite{tileup}.}
\end{figure}

The design of the mechanics for the on-board electronics is based on the concept of mini-drawers (MD) hosting each
up to 12 PMTs. A Main Board equipped with a Daughter Board is placed on one side of the MD structure,
the other side receives the HV distribution board (Figure~\ref{fig:meca}). 
For the Long Barrel 4 independent MDs are needed while 3 MDs plus 2 
mini-MDs are needed for the Extended Barrel. The concept was tested and validated in the November 2018 test beam campaign. 
The advantage of such a design relies in the modularity and the robustness
of the electronics as they are completely independent from one another. It also makes it easier for the maintenance as
one MD is a quarter of the length of a drawer in the current system.

\subsection{On-detector electronics}

\subsubsection*{The Front-End board}

Three different Front-End board (FEB) R\&D programs were pursued in parallel to identify a design with best performance for the HL-LHC, namely QIE~\cite{qie}, FATALIC~\cite{fatalic} and 3-in-1~\cite{3in1}.
The last, which is an evolution of the present system with more radiation hard components, was chosen as it is based on a very well known design.
These cards are being used in the Tile Demonstrator module (see Section~\ref{sec:demo}) for their compatibility with the current Level-1 trigger system
as they provide an analog trigger output.
Since the trigger in Phase-II will be fully digital, another version of this FEB - called FENICS - will no longer have an analog trigger output.
The PMT signal is shaped and amplified into two gains with 32:1 ratio and also into a slow integrator output.
The slow integrator readout measures the current due to p-p minimum bias interactions 
providing Luminosity Monitoring. It is also used to readout a Cs radiative source that can be moved trough the cells with a hydraulic system 
for calibration and monitoring of the PMTs gain and scintillator light yield.
The FEB features a 12-bit DAC and precise capacitors for the calibration of the electronics response using charge injection pulses.
The linearity of the system is under extensive studies and the dynamic range (typically about $0-25$~pC for the high gain and $25-800$~pC for the low gain) 
is under tuning to account for the usage of HV active dividers which improves the PMT gain by up to 20\%.

\subsubsection*{The Main Board}

\begin{figure}[htbp]
\centering 
\includegraphics[width=.45\textwidth]{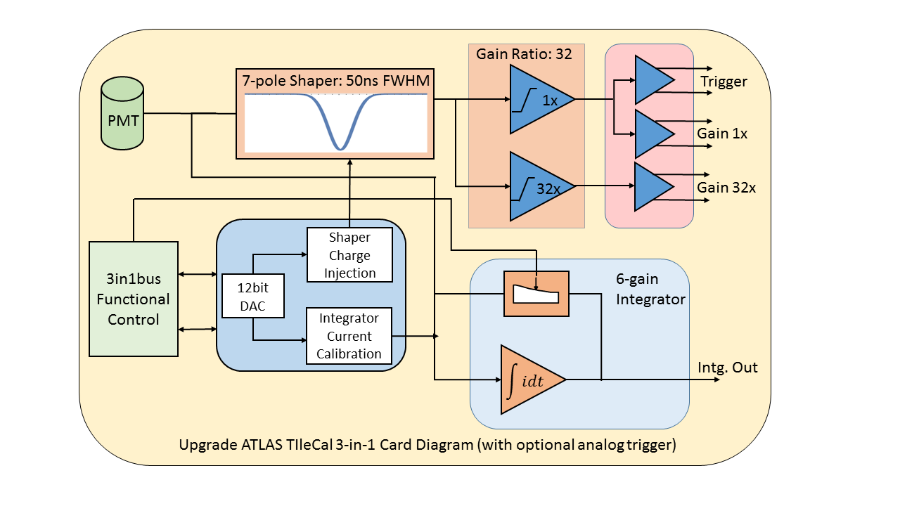}
\qquad
\includegraphics[width=.49\textwidth]{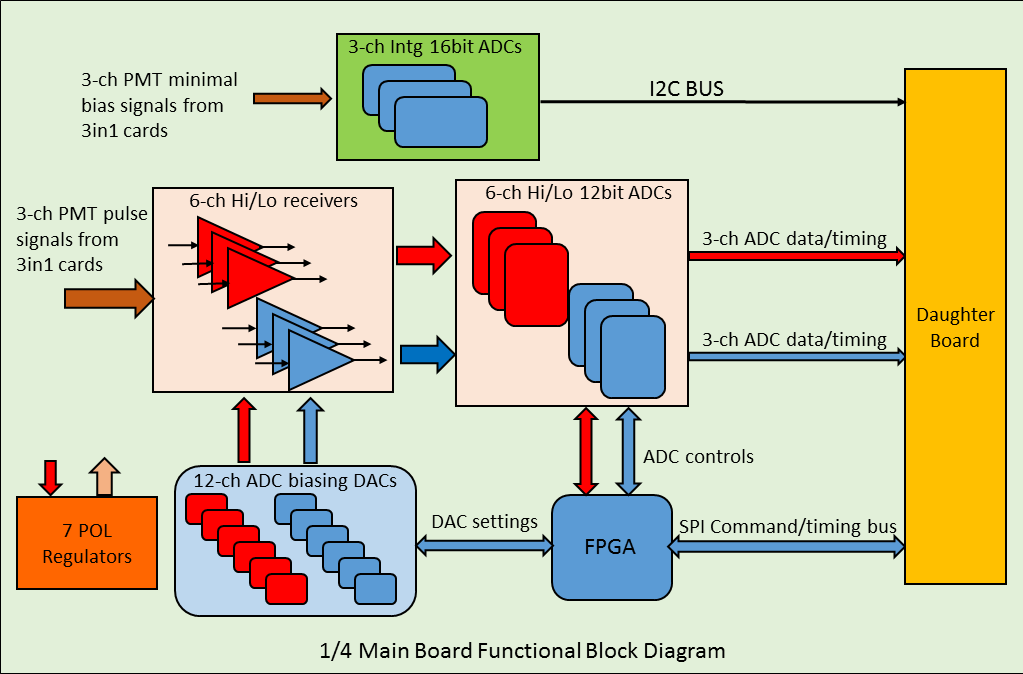}
\caption{\label{fig:mb-3in1} Functionality of the 3-in-1 card for Phase-II ({\it left}) and the Main Board ({\it right})~\cite{tileup}.}
\end{figure}

The Main Board (MB) controls, powers and acts as a digitizer interface between 12 FEBs and the Daughter Board.
It is composed of 4 sections with an Altera Cyclone IV FPGA in each where up to 3 PMTs could be connected. The readout of each section is done through 6 channels of
40 Msps 12-bit ADCs for high and low gain and 3 channels 50 kHz 16-bit ADCS for the integrator slow readout.
Digitized data are sent to the Daughter Board at 560 Mbps per ADC channel. 
The board itself is divided into two independent parts with Point-Of-Load (POL) regulators powered by a dedicated +10V bricks on the Low Voltage Power Supply. 
Figure~\ref{fig:mb-3in1} shows the functionality diagrams for 3-in-1 and MB. 

\subsubsection*{The Daughter Board}

The Daughter Board (DB) is the last element of the on-detector electronics and it ensures a high-speed link with the off-detector
electronics. It is responsible of the clock and command distribution to the on-detector electronics. Communication with the MB is done through
400 pins FMC connector.
Data from the MB are collected, formatted and transmitted through optical transceivers at 9.6 Gbps GBT links per side.
Each independent side of the DB could serve up to 6 FEB channels.

The DB version, v4, has two redundant Quad Small From-factor Pluggable (QSFP) connectors for the data transmission with the off-detector
electronics, two Xilinx Kintex-7 FPGAs and two GBTx chips. This version was used in the Demonstrator at the test beam.

In the latest version, v5, Xilink Kintex Ultrascale+ FPGA and 850 nm multi-mode Small Form-factor Pluggable (SFP) connectors were used. 
During irradiation tests this FPGA showed significant occurrence
of Single Event Latch-up (SEL). A new version, v6, is being produced, where the FPGA will be replaced back by the Kintex UltraScale
which showed more robustness against SEL. In addition an on-board over-current protection mechanism will be added, ensuring
a rapid power-cycling of the board in case of SEL.

\subsection{Off-detector electronics}

The off-detector electronics are hosted in four ATCA crates with each serving for the readout of a full detector partition.
One crate hosts 8 Tile PreProcessor (PPr) boards\footnote{A Tile PPr board is composed of 4 CPMs mounted on an ATCA carrier board.}, and 8 TDAQi 
(see next sections).

\subsubsection*{The Compact Processing Module}
\label{sec:cpm}
The Compact Processing Module (CPM) is a 14 layer PCB, 1.6 mm thick board for the readout of up to 8 MDs.
It is a high-speed interface with the DB for the data processing and handling, and it ensures the clock distribution
and the configuration of the MDs. It provides signal reconstruction and energy calibration for each PMT at the LHC frequency.
A first prototype has been produced and is now under test.

The Tile Demonstrator PPr~\cite{ppr}, a readout board with relatively reduced capabilities compared to the CPM,
was produced to operate the Tile Demonstrator module during test beam campaigns and during operation in ATLAS
prior to Phase-II. This board is capable of serving 4 MDs, receiving data from the Daughter Board at 40 MHz through four QSFP modules
requiring a bandwidth of 160 Gbps. The data are stored in circular pipeline memories and are sent out upon the reception of a Level-1 trigger Accept (L1A) signal.
The Demonstrator PPr was intensively used at the test beam and allowed valuable experience of the system to be gained.

\subsubsection*{The ATCA Carrier and the TDAQi}
\label{sec:atca}

A full-size carrier board was made and tested. It receives 
4 CPMs and one TDAQi Rear Transition Module (RTM) and acts as an interface between them. It also
carries on its board an IPMC for the power managing, a network switch mezzanine board and the TileCOM interface
for communicating with the rest of the elements.
The TDAQi processes the signals from the CPMs and calculates trigger objects from a group of cells
and sends them to the ATLAS trigger system. It also interfaces the data from the CPMs with the Front-End LInk eXchange (FELIX).
A first version of the TDAQi has been made and is now under tests.

\subsection{High Voltage and Low Voltage power supplies}

\begin{figure}[htbp]
\centering 
\includegraphics[width=.47\textwidth]{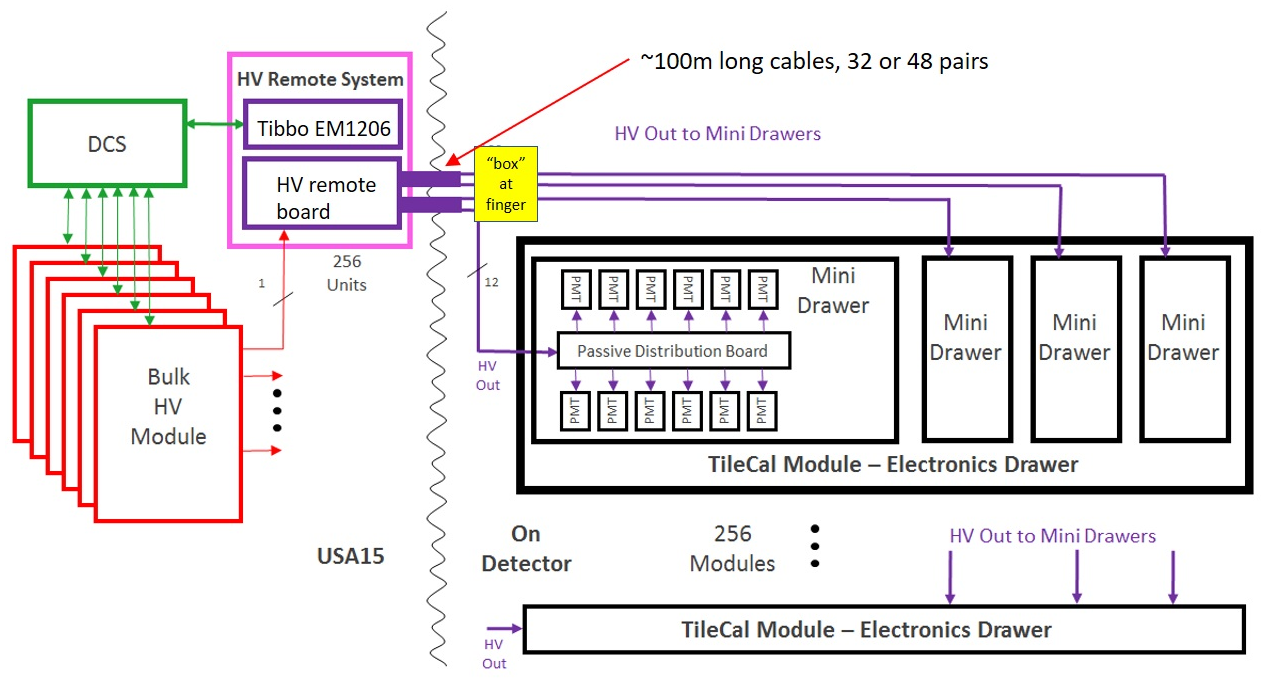}
\qquad
\includegraphics[width=.47\textwidth]{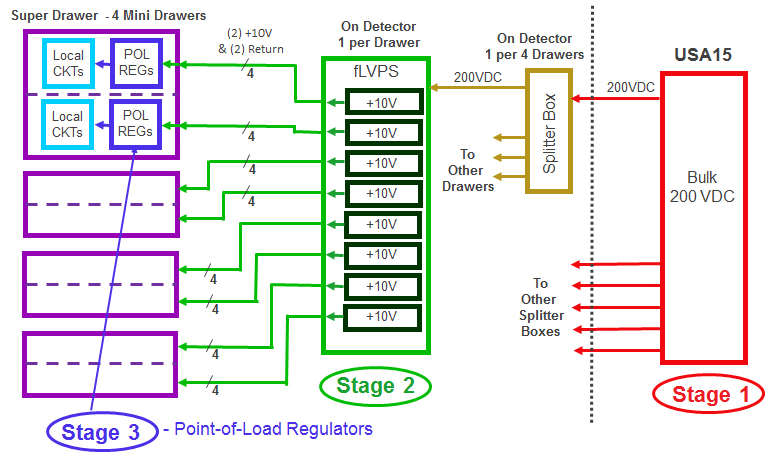}
\caption{\label{fig:hv-lv} Block diagram of the remote HV power distribution system ({\it left}) and of the 3-stage low-voltage power distribution system ({\it right})~\cite{tileup}.}
\end{figure}

For the High Voltage it was decided to adopt an off-detector HV distribution and regulation 
(HV remote) installed in the ATLAS service cavern (USA15).
The advantage of such a system is obviously that it isn't affected by the radiative ambiance 
of the detector cavern and that it is easily accessible for maintenance. Up to 48 pairs of wires
in 100 meter long HV cable are needed for each Tile module making a total of 256 HV cables for the 
whole detector. For the regulation system a 12-channel prototype was made and tested during the test beam campaigns. 
A full-size board with 48 channels was produced recently and is now under tests.
For the Low Voltage a three stage system based on the current Low Voltage Power Supply (LVPS) was chosen. 
First stage in USA15 provides 200VDC which are split to different DC/DC converters (bricks) of the second stage. 
These bricks converts the 200VDC to 10V inside the LVPS Box which carries also an ELMB Motherboard 
and a ELMB 2 board for the control and monitoring of the bricks. The 10V DC lines are then 
distributed 
to the power regulation system of the Main Boards. This system allows for a power redundant
distribution with 2 individual bricks per MD. In total 8 bricks are installed in each 
Tile LVPS Box. 
A long campaign of 6 irradiation tests and iterations has been performed in order to converge on a design 
that is radiation hard. Figure~\ref{fig:hv-lv} shows the block diagrams for the HV remote and the low voltage systems.

\section{Test beam and the Tile Demonstrator Project}
\label{sec:demo}
\begin{figure}[htbp]
\centering 
\includegraphics[width=.4\textwidth]{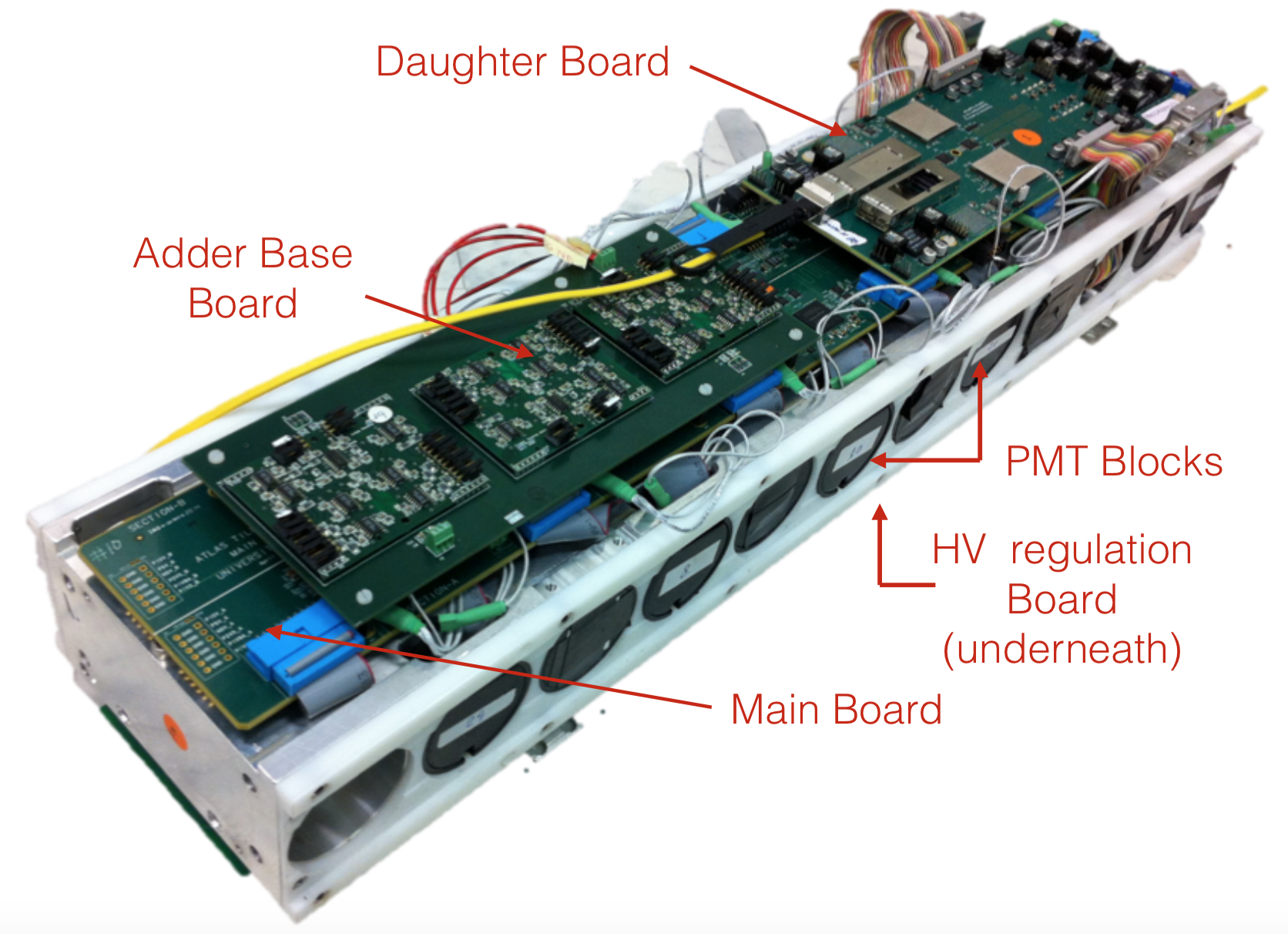}
\qquad
\includegraphics[width=.5\textwidth]{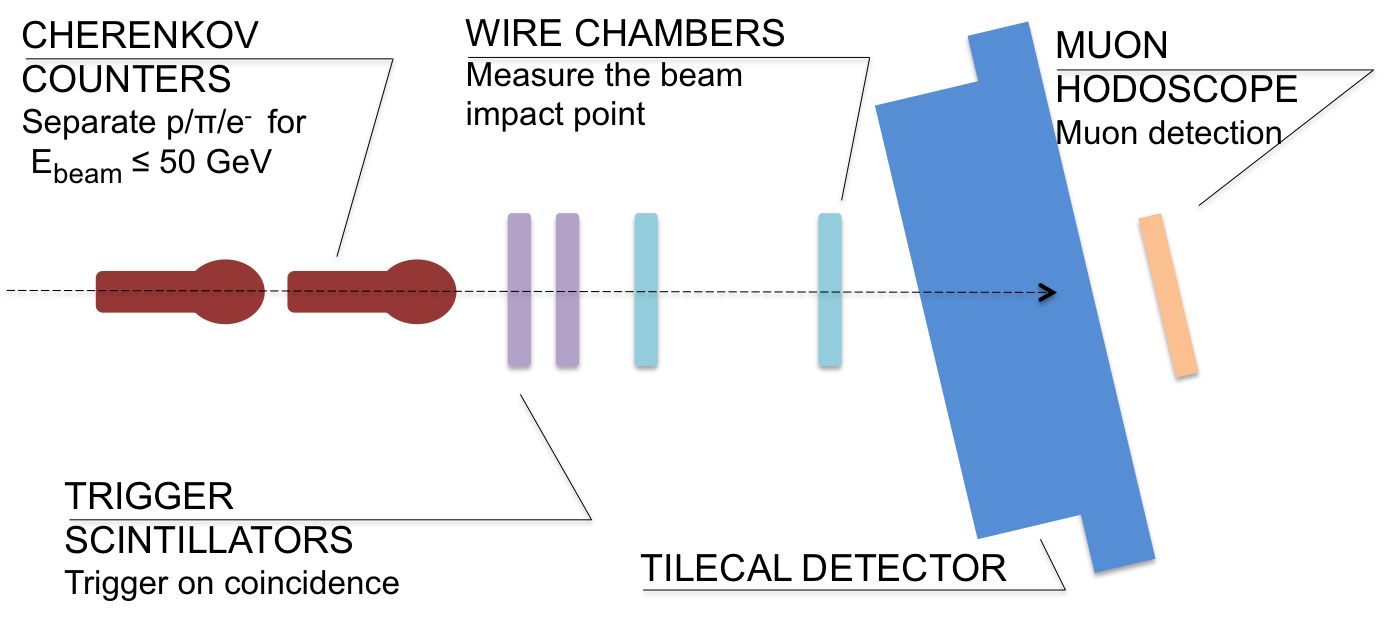}
\caption{\label{fig:demo-table} Demonstrator MD prototype ({\it left}) and schema of the test beam setup ({\it right})~\cite{tileup}. }
\end{figure}

Several test beam campaigns have been held at the H8 beam facility at the CERN SPS North Area.
Three Tile modules were stacked on a movable table. On the top, an Extended Barrel module instrumented
with three 3 MDs and 2 mini-MDs. The MDs were equipped with FENICS 
cards and Phase-II Main-Boards and Daughter Boards. On the middle, half of the 
Long Barrel module was equipped with the Tile Demonstrator electronics (4 MDs). This hybrid version carries 
on the present 3-in-1 cards instead of FENICS cards. This is because this module is meant to be 
compatible with the current ATLAS Trigger system allowing at the same time to test the other
Phase-II electronic boards. Two MDs were operated using the HV remote system while 
the other two were operated using the current on-board HV distribution through HV Opto boards.
The remaining modules were equipped, depending on the programs of the test beam,
with a combination of legacy drawers and MDs with different Front-End options.
The aim was to study the performance of the diffferent electronic options and to get a direct comparison with the legacs system. 
On the up-stream, Cherenkov Counters, Trigger Scintillators and Wire chambers are placed in order to
provide particle identification, trigger and position measurements (Figure~\ref{fig:demo-table}). 

\begin{figure}[htbp]
\centering 
\includegraphics[width=.55\textwidth]{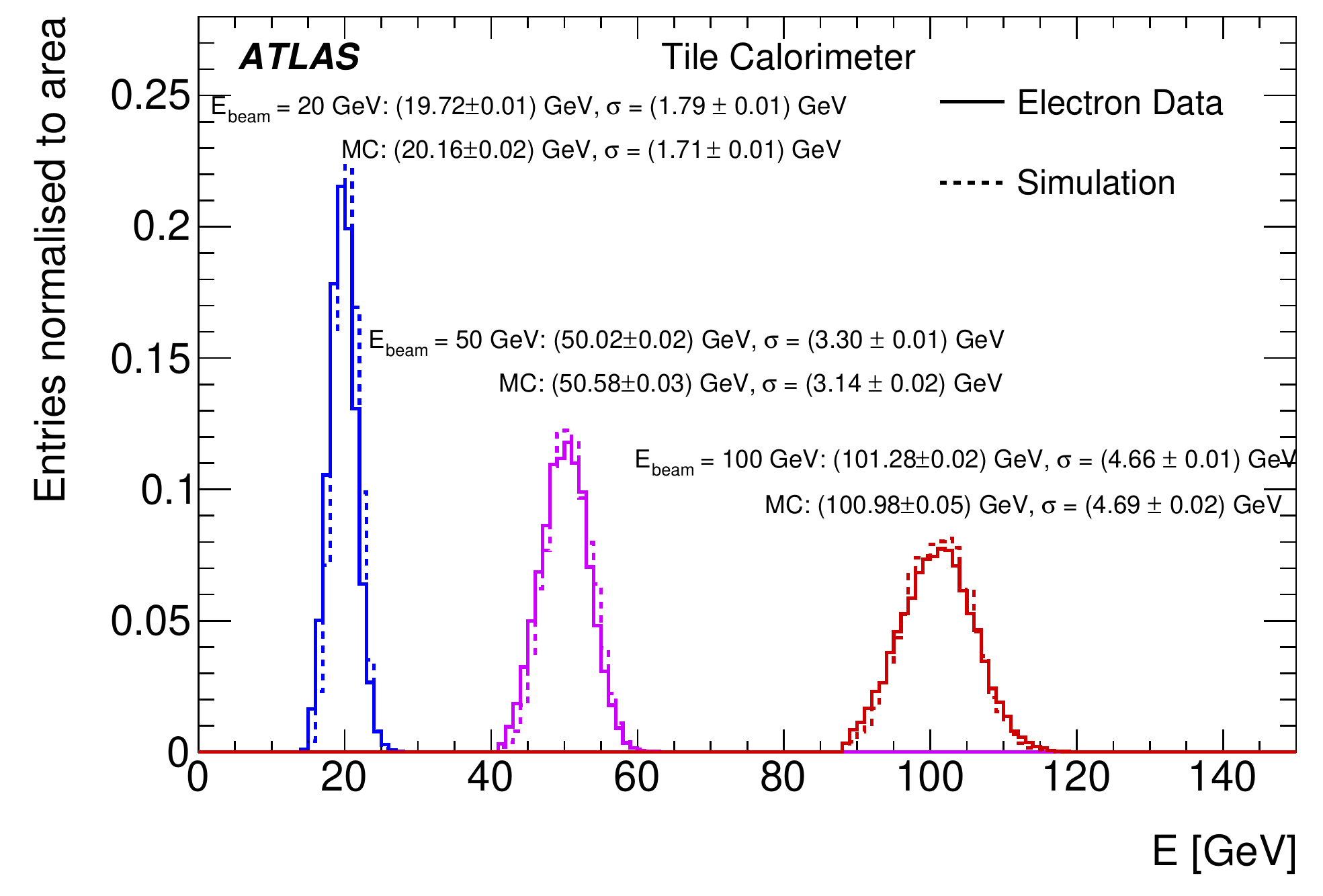}
\qquad
\includegraphics[width=.38\textwidth]{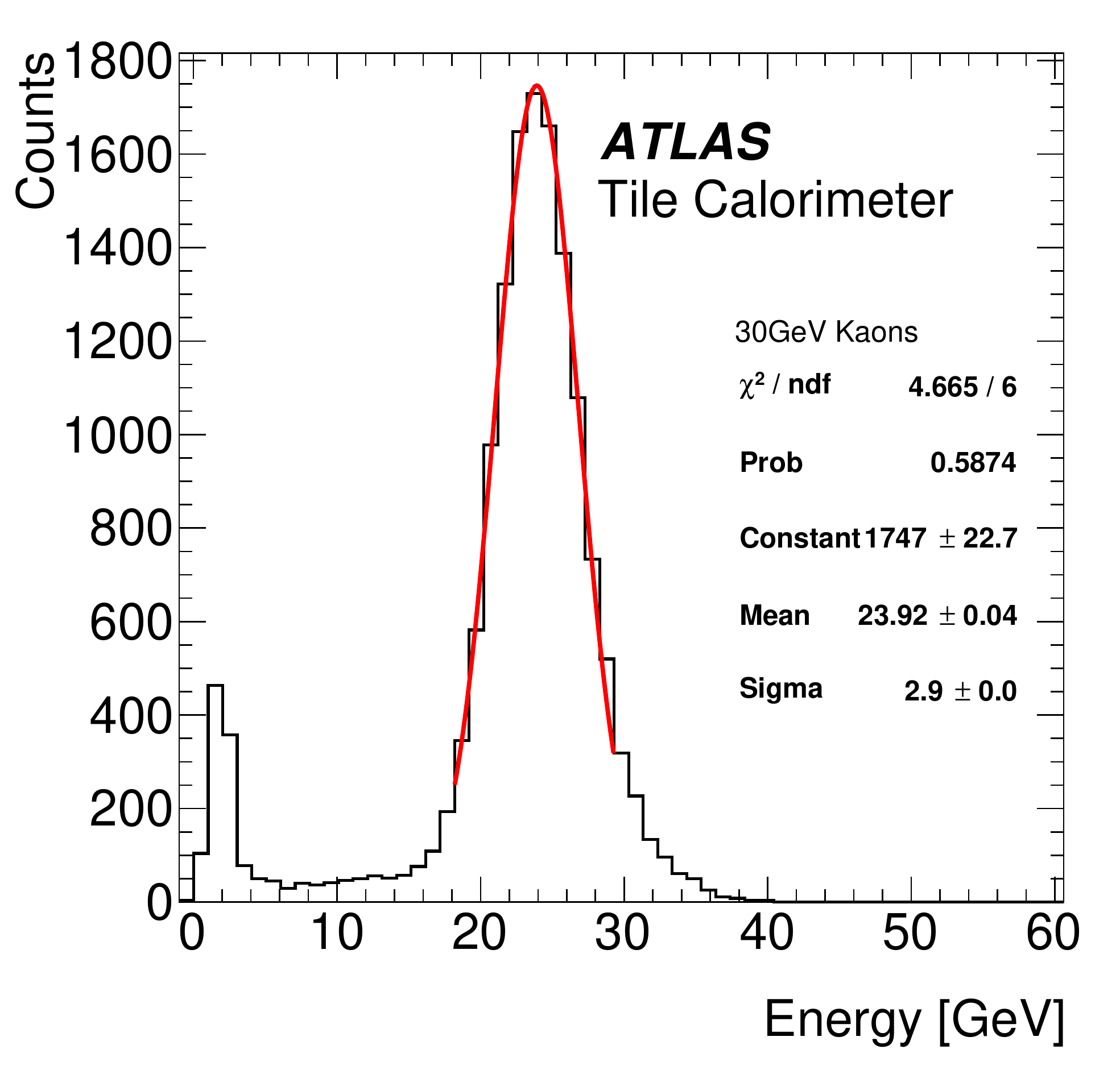}
\caption{\label{fig:tbplots} Distributions of the total energy deposited in the calorimeter obtained using electrons beams at different energies ({\it left}) and of the measured energy in the case of 30 GeV kaons ({\it right})~\cite{tileup}.}
\end{figure}

The modules with new electronics were integrated in the ATLAS TDAQ software and Detector Control 
System (DCS). The on-detector electronics were configured through the PPr and physics and calibration 
data were taken through the current ROD system as well as the FELIX system.
The good performance of the new electronics was demonstrated during these test beam campaigns.
Figure~\ref{fig:tbplots} demonstrates some of the obtained results.

After the replacement of some electronic boards with newer versions and the upgrade of the HV to be 
fully compatible with the remote system, the Tile Demonstrator was inserted in ATLAS on July. 
The aim is to operate it together with the legacy system during the Long Shutdown 2, 
and possibly during Run 3.
During the insertion, some mechanical interference between the mini-drawers and the girder were 
discovered. This leads to small modifications on the metallic parts of the mini-drawers.
The Demonstrator has operated since then through the TDAQ and the ATLAS DCS 
system. Stable performance in terms of low noise and good CIS and laser signals were
demonstrated. The analog trigger interface to the ATLAS trigger system is under test. First tests using 
laser signals showed promising results in most TileCal trigger towers. 

\section{Conclusion}

A wide R\&D program to redesign the on-detector and off-detector electronics of the ATLAS TileCal 
for the HL-LHC is progressing well. All elements of the full readout 
chain are in an advanced state and prototypes have been produced and tested extensively.

Most of the radiation sensitive components have been 
certified. The performance of the new electronics has been evaluated throughout
multiple test beam campaigns from 2015 to 2018. The results are used to tune the latest 
versions of the electronics.
The Tile demonstrator module was successfully integrated in the current system.

\end{document}